\documentclass[12pt]{revtex4}
\input epsf
\newcommand{\beq}{\begin{equation}}
\newcommand{\eeq}{\end{equation}}
\newcommand{\beqa}{\begin{eqnarray}}
\newcommand{\eeqa}{\end{eqnarray}}

\begin{document}

\title{\large Some Comments on the Putative $\Theta^
+$ (1543) Exotic State}
\author{S. Nussinov }
\email{nussinov@post.tau.ac.il}
\affiliation{Department of Physics and Astronomy\\
University of South Carolina\\
Columbia, Sc 29208}
\altaffiliation{On sabbatical leave from 
Tel-Aviv University, Tel-Aviv, Israel}

\date{\today}

\begin{abstract}

  We point out that existing $K^+d$ scattering data available in the
  PDG ( Particle Data Group compilation )
  suggest some fluctuations in those momentum bins where the (Fermi
  motion broadened) $\Theta^+$[1543] resonance recently indicated in
  many gamma nuclear reactions and predicted six years ago by Diakonv
  Petrov and Polyakov might have shown up.
   The $I=0$, $J^P=\frac{1}{2}^+$ P-wave channel should have a universal peak
  cross section of $\sim 37$ mb at resonance.
   The smallness of the effect seen in $K^+d$ with the $\delta \sigma $
 fluctuations being  less than 4 mb imply an indirect bound
 $\Gamma_{\Theta^+} < 6$ MeV, far stronger than the direct gamma-d measurements.
  This renders the theoretical interpretation of the new state very
 difficult.

\end{abstract}

\maketitle

\section{ Introduction}

   Indications for a $K^{+}n$ resonance  at a mass of 1543 MeV were found in several  Photon deuteron collision experiments \cite{spring8,Class}  the final $K^{+}K^-pn$ state  and also in $K^{+} - Xe$ collisions\cite{ Lena} suggest that a low-lying narrow  5-quark $\bar{s}uudd$ state, the $Z^+$ exists.  Capstick, Page and Roberts (CPR) \cite{CPR} suggested that $Z^+$ is an isotensor.   Such a state can be produced in $\gamma+d \rightarrow nK^{+}K^-p$ reactions   but decays slowly into  $I=0$ or $I=1$ final $K N $ states due to the I-spin violation required, explaining the narrow width.

  In a remarkable 1997 paper, Diakonov, Petrov and Polyakov (DPP)\cite{DPP}
  started with an SU(3) extension\cite{MN} of the Skyrme model and predicted a low-lying SU(3) ${\mathbf{\overline{10}}}$ anti-decuplet of $\frac{1}{2}^{+}$ baryons with $\Theta^+=\bar{s}uudd$
  serving as its $I=0$ lowest hypercharge and lowest mass entry.
  Identifying the $I=1/2$ doublet in this antidecuplet with the
  N(1710) state they predicted $m_{\Theta^+} \sim 1530$ MeV, close to the experimental 1543 MeV
  value. Having $I=0$ it can appear in $K^{+}n$  and $K^0p$ but not in
 the well studied $K^{+}p$ channel, which indeed has no resonances!
  Also DPP estimated a width $ \Gamma_{\Theta^+}\simeq 15 MeV$,  consistent  with the present direct experimental upper bounds $\Gamma < 20$ MeV. 
   Unfortunately, the starting point of DPP, namely the baryonic soliton in
 large $N_c$ and chiral SU(3), makes the paper somewhat inaccesible.

  In the following we present  naive quark model (NQM) arguments, which explain
  in simple terms why the $\bar{s}uudd$, $ I=0$, $J^P=\frac{1}{2}^{+}$ state is likely to be
  low-lying. The argument clarifies connections to earlier Bag-NQM predictions of
  Hexa-\cite{Jaffe1},Penta-\cite{Lipkin1,Richard} and Tetra- quarks\cite{Manohar,Gelman}. It also implies that $I(\Theta^+)=2$, the simplest   explanation for the narrow width, is extremely unlikely.

   Our main observation which we present next is that the \emph{lack} of a prominent
$\Theta^+$ signature in $K^{+}-d$ collisions restricts the width to
$\Gamma_{\Theta^+}<6 $ MeV making $I(\Theta^+) =0$ barely consistent even in the special
 context of DPP's anti-decuplet.

\section{Bounds on the $\Theta^+$ width}

\noindent We use the following  simple observations:

\noindent{\textbf{(i)}} The $l=1$ orbital angular momentum of the K-N $\Theta^+$ resonance predicted by DPP is almost model-independent: In $l=0$ the attractive nuclear forces cannot yield narrow
resonances but only threshold enhancements as in S-wave N-N scattering
and $l> 1$ is unlikely for the low-lying $\Theta^+$.

\noindent{\textbf{(ii)}} The general expression for the total $K^{+}-n$ cross section is:
   \begin{equation}
\sigma_{K^{+} n}(p) = {4 \pi \over p^2} \Sigma_{l}(2l+1)\ E \ \sin^2{\delta_{l}(p)
}
\end{equation}
with $p$ the momentum of K or n in the center of mass Lorentz frame, and
  $E = E(I,J)$ reflects the I spin and angular momentum projection "Clebsches".
  The resonant phase shift is $\delta_l(p)= \pi/2$ and the relevant partial wave
  cross section saturates at a universal value. For the $\Theta^+$ with $p=0.27 $ GeV
 and $l=1$ the value is:
 \begin{equation}
\left. \sigma_{ \Theta^+ }\right|_{res} \simeq {4 \pi \over p^2} \cdot 3 \cdot \left(\frac{1}{2}\right)\cdot\left(\frac{1}{3}\right) \simeq 37 \ \ \mathrm{mb}
\end{equation}
The factor $ 1/2$ reflects the equal-magnitude projections of the
  $A(K^{+}-n )$ and $A(K^0-p)$ scattering amplitudes on the $I=0$ resonant
 amplitude and the $I=1$ non-resonant amplitudes (implying also equal
 elastic $K^{+}n \rightarrow K^{+}n$ and charge exchange (CEX) $K^{+}n \rightarrow K^0+p$ parts ).  The second factor of $1/3$ is the probability that the $l=1$ orbital and
  $s=1/2$ spin of the nucleon add up to the desired $J=1/2$.

 \noindent{\textbf{(iii)}} The implications of the universal resonant cross section in specific
 cases depend on the width of the relevant state.  Thus the minute neutrino electron $\bar{\nu}_e-e$ cross section jumps by $\sim 10$
 orders of magnitude at the $\pi^{-}$ mass. However, the tiny width
 $\Gamma_{\pi^{-}}\sim 10^{-17}$ GeV makes this unobservable.

Let the width of the putative new resonanc be $\Gamma_{\Theta^+} = g \ 20 \mathrm{MeV}$.
The 20 MeV "reference" value predicted by DPP, is consistent with direct
 observations.
 If neutron targets and/or monochromatic $K^0$ beams were available the $\Theta^+$
 would manifest as a threefold increase in the  $K-N$ cross section
 if $ \Gamma_{\Theta^+}$ exceeds the step size in the scan.
In reality neither is available. However, as we show in the next section,
 existing $K^{+}$-deuteron data indirectly exclude widths larger than $ \sim 6$ MeV.

 \noindent{\textbf{(iv)}} The $KN$ channels have been extensively studied in the 60's using ,
 in particular, bubble chambers. The absence of resonances in $K^{+}p$, dramatically contrasted the resonances or $\bar{K}N$  ``bound states" (decaying to $\Lambda + \pi$) found in $K^-p$ scattering.

  The PDG\cite{PDG} plots of $K^-N$ cross sections indicate, for CMS energies below
 $m_{\Theta^+}=1540$ MeV  huge $O(80)$ mb cross sections, which , as expected for
 the exothermic process, blow up like $1/v$ towards threshold.

  Using $K^0$ beams (obtained by$K^{+}$ CEX reactions) to search for the $\Theta^+
$ in $K^0p$ reactions in Hydrogen bubble chambers is inherently impossible.
 Despite some material, ``index of refraction" effects, the propagating states
 are, as in vacuum, almost an equal mixture of $K_{\mathrm{short}}$ and $K_{\mathrm{long}}$ with the mildly relativistic $K_S$'s decaying into two pions along paths of order 1-2 cm,
 shorter than the mean free paths for K-N interactions. The remaining $K_L$'s
 are an equal mixture of $K^0$ and $\bar{K}^0$, and the huge cross section
 of the $\bar{K}^0$ components can mask the fine structure in the $K^0-p$
cross section which is being searched.

This does  not hinder (and even helps!) searching the $\Theta^+$ in $K^{+}d$ data.
 The total $K^{+}d$ cross section reflects the $\Theta^+$ resonance in
 the $K^{+}n$ channel albeit broadened by the neutrons' Fermi motion.
 To estimate the latter we use a Yukawa $  \exp{-\mu r}/r $ deuterons' wave
 function with:
         \begin{equation}
\mu=\sqrt{m(N).BE} \sim 50 \mathrm{MeV} \, .
\end{equation}
The probability $P(k)$ that the neutron has a "Fermi" momentum $k$ is the
square of the Fourier transform of the wave function
\begin{equation}
 P(k) \propto { 1 \over ({\mu^2+k^2})^2 }\, .
\end{equation}
If the lab momentum of the $K^{+}$ in the rest frame of the deuteron is $q$,
 then for any given $|k|$ the ``lab" momentum for the $K^{+} n$ collision
 of interest is uniformly distributed in the $(q-|k|,q+|k|)$ interval
 This broadens the putative $\Theta^+$ resonance plotted versus the
 $K^{+}d$ lab momentum by  $\Delta q = 2 \mu$, i.e by 100 MeV.
The mapping of W(cm) to $q$-lab momentum in the vicinity of $W=m_{\Theta^+}=1543$ MeV
 doubles the original $\Gamma_{\Theta^+}= g \ 20$ MeV to $ g\ 40$ MeV in the $q$ variable.
 Fermi motion broadens this into $ (g\ 40 +100) $ MeV, i.e., by a factor
 \begin{equation}
f={g\ 40+100 \over g\ 40}\, ,
\end{equation}
 and in the process ``dilutes" the peak resonance cross section by $1/f$ to
 \begin{equation}\label{deltasigma}
 \left. \delta \sigma \right|_{\Theta^+ \mathrm{res}}= { 37 \over 1+2.5/g}\ \mathrm{mb}
\end{equation}

 A careful examination of the total $K^{+}d$ cross section plots reveals that in
 the 200 MeV interval of lab momenta 0.5 - 0.7 GeV/c which corresponds to
 W=1.5-1.6 GeV there are intriguing fluctuations in all experiments!
 These $\delta \sigma_{K^+d}=2-4$ mb fluctuations, if confirmed by detailed
 analysis would constitute independent evidence for $\Theta^+ \simeq (1543)$ MeV.
 If these fluctuations are indeed due to the $\Theta^+$ resonance we infer from
Eq. (\ref{deltasigma}) and the conservative estimate $\delta (\sigma) < 4$ mb, that $g < 0.3$ and
\begin{equation}
 \Gamma_{\Theta^+} < 6 \ \mathrm{MeV}\, .
\end{equation}

   The above upper bound on the width of a putative $K^{+}-n$ $\Theta^+$ resonance
relies only on the total  $Kd$ cross sections and may be highly conservative.
 Half the resonant cross section is CEX which dramatically manifests via the
 pions from the $K_S$ decay, and absorption of $\bar{K}^0$'s regenerated via
 the $K_L$'s.
A direct study of the viability of the $\Theta^+$ in the specific experiments
 used in the PDG is beyond the scope of this paper. It may imply even
 smaller $\Gamma_{\Theta^+}$, or, more dramatically, verify the existence of the $\Theta^+$!
 
\section{ Can we have a narrow $\Theta^+$ ? }

 Is our upper bound $\Gamma_{\Theta^+}< 6$ MeV consistent with $I_{\Theta^+}=0$? 
 CPR suggested $I_{\Theta^+}=2$ since they expected an unsuppressed ``fall-apart"
width of $\Theta^+$ of several hundred MeV.
 Let's compare $\Gamma_\Delta$, the widths of the $(3,3)\ 1230$ P-wave pion
-nucleon
 resonance with $p(3,3)=0.27$ GeV/c and that of the P-wave $\Theta^+$ with a cm frame
 momentum $p' =.23$ GeV/c with a $\sim 40\%$ stronger centrifugal barrier suppression
 of the decay rate. The decay of the $\Delta$ but not that
 of $\Theta^+$ requires creating an extra light $d\bar{d}$ quark pair whereas any
break-up  of $uudd\bar{s}$, into $\bar{s}u + udd = K^{+}n$ or $\bar{s}d+ duu = K^0p$ is
allowed.
In the $1/N_c$ expansion\cite{Witten} or other non-perturbative frameworks\cite{CNN}- this should
  suppress  the $\Delta$ width relative to that of $\Theta^+$,  suggesting
\begin{equation}
\Gamma_{\Theta^+} > .6\  \Gamma_{ \Delta(3/2,3/2)}= 70\ \mathrm{MeV}\, ,
\end{equation}
apparently excluding $I(\Theta^+)=0$. 

CPR suggest that the widths of \emph{all} the anti-decuplet states of DPP are larger
 than the DPP estimates. We note, however, the remarkable internal consistency of the DPP scheme. Thus, following DPP, let $N(1710)$, $J^P=\frac{1}{2}^+$ be the $I=1/2$ member of $\mathbf{\bar{10}}$, e.g., 
\begin{equation}
 N^+={2 \over \sqrt{3}} \bar{s}s\, uud + {1 \over \sqrt{3}} \bar{d}d\, uud \, .
\end{equation}
It has a relatively PDG small width $\Gamma_{N(1710)}=100$ MeV with only 10-20 MeV
 partial decay width into the $N\pi$ channel. If indeed the $N(1710)$ and
 the $\Theta^+$ have similar dynamics then   $\Gamma_{N(1710)\rightarrow N \pi}=10-20$ MeV implies, as we shall shortly explain,  $\Gamma_{\Theta^+\rightarrow KN}= 3-6$ MeV, which
 is consistent even with the more stringent upper bound that
  we inferred from the  absence of prominent bumps in the $K^{+} d$ data.
 
 The probability that the $N(1710)$ does not contain any $s\bar{s}$ and can
 decay without ``Zweig-Rule" violation into a non-strange $N\pi$ final state is
  only 1/3. Along with the centrifugal barriers ratio,
   this yields the above estimate for $\Gamma_{\Theta^+}$.

\section{Why a Low-Lying $\Theta^+$ is Likely to Have I=0 ?}

  If further analysis along the lines of sec II above will imply a $\Theta^+$
  much smaller width than the present 6 MeV upper bound, then $I(\Theta^+)=2$ may be
 imperative.  The following explains why a low lying isotensor state is (theoretically)
unlikely:

  Let us first present the argument at the hadronic level.
 The lowest hadronic channel accommodating an isotensor $\Theta^+$ is $\Delta(0) K^{+}$
 with a threshold of $1230 + 495 = 1735$ MeV $\sim 300$ MeV higher than the KN
 channel in which the $I=0$ (and $I=1$) states can appear.
 In general channels with lower I spins have stronger binding. Thus in the
 $\bar{K}N$ channel, the $I=0,1$ $\Lambda$ and $\Sigma$ are the lowest $J^P=\frac{1}{2}^+$
 bound states with $\Sigma-\Lambda = 80$ MeV. This can be  explained  as being
 due to the exchange of the vector $\rho$ meson . If the latter couples to
 isospin then the exchange generates in general an interaction potential
 $V \sim \vec{I}_a \cdot \vec{I}_b$ in a channel a,b. Treating this perturbatively yields, when
 $I_a$ and $I_b $ add up to a total $I$  a contribution:
  \begin{equation}
  \delta \rho_{ a,b}= c \left[ I(I+1)-I_a(I_a+1)-I_b(I_b+1) \right]
\end{equation}

  Comparing the $I=2$ and $I=1$ $\Delta(3/2,3/2) K$ state and in turn the
 $I=1$ and $I=0$ NK states and assuming for simplicity that the same factor $c$ appears
 in both, and also in the $\bar{K}N$ channels we find that a putative
  isotensor $\Theta^+$ will be less bound than the I=0 $\Theta^+$ by 200 MeV.
  The isotensor $\Theta^+$ in the $\Delta K$ channel is thus expected to be
 ~ 500 MeV heavier than the corresponding $I=0$ KN state!
  
  The pattern of binding in hadronic channels via OBE (One boson exchange)
 potentials can often be related to the underlying quark structure of the
 hadrons involved. Thus much of the hard core N-N potential reflecting a
repulsive vectorial $\omega$ exchange with $\omega$ coupling to the total
number of light u,d quarks seems to trace back to the Pauli exclusion
 principle. Also attractive $\rho$  exchange could reflect the possible amelioration of this for smaller numbers of equal flavor light quarks in the two hadrons say in n-p versus nn or pp (
which  indeed have I=1). 

Both the solitonic Skyrme picture used by DPP to motivate the $\Theta^+$, and   earlier Bag model calculations suggesting various Hexa- ,Penta-, and
Tetraquark ``exotic" states all treat the system as a ``single bag" or a ``single, connected
 color network" rather than as two separate color singlet hadrons.

  In a Naive Quark Model the non-exotic $qqq$ baryons and $q\bar{q}$
mesons are viewed as  bound states of mildly relativistic ``constituent quarks".
 The latter ``quasi-particles " are made of bare quarks with the relevant
 flavor and  gluon+$q\bar{q}$ clouds, have effective masses $m_u \sim m_d \sim$ 350 MeV and $m_s \sim$ 450-500 MeV. 
Once quark confinement is imposed via the ``Bag", or, via confining
 potentials, the most important ``residual" force between the (constituent)
 quarks is the ``hyperfine" chromomagnetic interaction contributing
  \begin{equation}\label{hyperfine}
\Delta M = C \sum_{i,j} \frac{1}{m_i m_j}(\vec{S}_i \cdot \vec{S}_j) \left(\vec{\lambda}_i \cdot \vec{\lambda}_j \right) 
\end{equation}
with $C$ setting the overall scale of the interaction matrix element, 
$\vec{S}_i, \vec{\lambda}_i,$ and $m_i$ being the spin, color matrix and mass of the quark
(or anti-quark) $q_i$. With hadron-independent $C$ this explains the observed
pattern:
 $\rho-\pi = 650$ MeV  approximately equal to twice  $ \Delta-N \simeq 300$ MeV, and approximately equal to three halves of $K^*-K = 400$ MeV, which is equal to twice $\Sigma(1380)-
 \Sigma = 200$ MeV.
 This is due to 
\begin{equation}
 \langle \vec{\lambda}_i.\vec{\lambda}_j\rangle _{\mathrm{Baryon}}= \frac{1}{2} \langle\vec{\lambda}_i.\overline{\vec{\lambda}_j}\rangle_{\mathrm{Meson}}
\end{equation}
since in the baryon each di-quark pair is, by overall color neutrality,
 in a $\mathbf{\bar{3}}$ SU(3) color representation whereas the $\mathbf{3}$ and $\mathbf{\bar{3}}$ in the
 meson add up to a singlet.  In addition, we have the inverse mass factors
\begin{equation}
 {1 \over m_s m_q } \sim \left(\frac{2}{3}\right){1 \over m_q m_{q'}}\, .
\end{equation}
 where $q$,$q'$ refer to the lightest $u$, $d$ quarks.

 The idea that also the pattern of the lowest-lying multi-quark states can
 be inferred by minimizing the overall hyperfine interaction energy has been
suggested early on (in a bag model context), by R. Jaffe\cite{Jaffe2} and developed by
 several authors \cite{Lipkin1, Richard}.
 It suggested that the low-lying ,$0^+K\bar{K}$ threshold states with $I=0$
 and $I=1$, namely $f(980)$ and $a(980)$ respectively, are such $s\bar{s}q\bar{q}$ states. Many other
four- and five-quark states and in particular the new BaBar $D[s](2317)$ state
 could have a similar origin \cite{N,Lipkin2}. 
 
 Jaffe further suggested \cite{Jaffe1} that the di-baryon $H =uuddss$ lowest lying $O^+$
 state is particularly strongly bound: $m_H< 2 m_\Lambda$ and the $H$ decays
only via repeated weak interactions. While dedicated searches did not find
a stable $H$, stronger binding in these flavor and quantum numbers is likely.
 Later on, similar considerations  suggested that a stable $\bar{c}suud$
 pentaquark is even more likely.

  The remarkable success of the NQM or bag models is from the point
 of view of a QCD purist, rather surprising. 
  Equally surprising is the fact that the Skyrme model originally embedded
 in the large $N_c$ Chiral SU(2) and extended to SU(3) \cite{MN} reproduces the
 same pattern of octet and decuplet ground state  baryons and \cite{Bala}
 even indicates the special $H$ state in the di-baryon extension of the Skyrme model.
 This remarkable feature persists also for the new $uudd\bar{s}$ state in the
 anti-decuplet that DPP derive as the next Skyrme model level.
 The following brief discussion qualitatively motivates the above statement.

 To see most clearly the physics involved consider first an idealized setting
 were the ``heavy" $\bar{s}$ in $\Theta^+$ is replaced by a much heavier $\bar{Q}$.
 The heavy (anti-) quark ``sits" at the origin serving as a static $\mathbf{\bar{3}}$ SU(3)
 color source\cite{NW} and the H.F $\bar{Q}-q_i$ interactions which are
 proportional to $1/m(Q)$ can be neglected.
 The invariance under overall rotations of the light quark system with respect
 to the heavy quark ``anchoring point" implies  a vanishing relative angular
 momentum $L(qqqq, \, \bar{Q})=0$.

 The $J^P$ and isospin of the lowest $\bar{Q}u_1 u_2 d_1 d_2$ state are fixed by
 minimizing the total energy of the $qqqq$ system , which we assume is
 dominated  by the chromomagnetic H.F Interaction of  Eq. (\ref{hyperfine}).
  Color neutrality of the pentaquark implies that $qqqq$ is in the fundamental
$\mathbf{3}$ representation:
$ \lambda_{u_1}+\lambda_{u_2}+\lambda_{d_1}+\lambda_{d_2} = \mathbf{3}$
  and we have anti-symmetry under the joint exchange of the color,spin,
 flavor and orbital parameters of any $q_i$ and $q_j$.

  For a first, heuristic, go-around we make use of the diquark concept.
 Any $xud$ system like $ \Lambda_s/ \Sigma_s$ , $\Lambda_c/ \Sigma_c$,
 etc. contains a diquark made of the light $ud$ system with overall $\mathbf{\bar{3}}$
 color. The H.F interaction makes the $I=O$, $S=0$ $ud$ and corresponding $\Lambda$'s -
 more tightly bound then the $I=1$ $ud$ and corresponding $\Sigma$'s. The effect is
 stronger for higher $m(x)$ and weaker $xq$ H.F interactions (the latter prefer
 triplet $ud$ so that $x$ can anti-align with both $u$ and $d$).
 By extrapolating from the $s$-$c$ quarks we expect $m(ud,S=1,I=1) - m(ud,S=0,I=0) \sim
200$ MeV.  Since the colors of the two quarks have to add up to a $\mathbf{\bar{3}}$ the $ud$ system  is anti-symmetric in color and hence, by overall Fermi statistics we have the
 $I=S$ ``locking" for the S-wave quarks.

 Consider then a particular pair of $ud$ diquarks , say $u_1 d_1$ in $\mathbf{\bar{3}}$ of
 color and $u_2 d_2$ in a $\mathbf{\bar{3}}$ of color. The overall $qqqq$ system will be lighter by  400 MeV(!), if both diquarks are in the $I=0$, so that $I(qqqq)= I(\Theta^+)=0$
 as compared with the case when both diquarks are in $I=1$ and $I(qqqq)=I(\Theta^+)=2$.
 In the energetically-favored case we also have vanishing total spin: $S(qqqq)=0$.
 An overall $I=S=0$ $ qqqq$ system is consistent with having the lower lying
 configuration in the other pairing: $   I(u_1 d_2)=S(u_1 d_2) = I(u_2 d_1)=S(u_2 d_1)=0$
 as well.
 
 The two $I=S=0$ diquarks are effectively identical bosons and the wave-function
 should be symmetric under the exchange of the diquarks. However the color
coupling  of the two $\mathbf{\bar{3}}$'s of the two diquarks must be anti-symmetric to ensure that $qqqq$  be in the fundamental $\mathbf{3}$ representation of color. Hence, the two diquarks
 must be in a relative orbital angular momentum $l=1$ and hence $j(qqqq)=1$. Upon adding this to
 $s(\bar{Q})=1/2$ the total $J=J(\Theta^+)=1/2$ is favored by the L-S coupling.
 Also the parity of the system will be positive on account of the intrinsic negative
 parity of the $\bar{Q}$.
 
 The above NQM finger-counting/hand-waving is admittedly crude. We need to
 allow the $q_i q_j$ pairs to couple also to color sextets, and for
 $\bar{Q}= \bar{s}$ to carefully consider the H.F interactions of the $\bar{s}$
   as well. Indeed, these extra H.F. interactions may  modify the energetics. Yet these
considerations
 exclude the $I=2$ assignment and select precisely the DPP $\Theta^+$ state, which
 emerged there in the framework of the SU(3) extended Skyrme model! We cannot
 predict $m_{\Theta^+}$ but neither can DPP without ``anchoring" to the $N(1710)$.

 \section{ Summary and Conclusions }

  In this note we have greatly praised the DPP paper and tried to
reproduce its predicted $I=0$, $J^P= \frac{1}{2}^+ \Theta^+$ using simpler arguments.
 We also noted  some curious and extremely intriguing
fluctuations in the $K^{+}$-deuteron total cross section data compiled in the
PDG as a function of the $K^+$ lab momenta precisely in those bins where
the Fermi- motion broadened $\Theta^+$ state is expected to show up.
 These could clearly establish the existence of $\Theta^+$  if indeed confirmed by a more careful
analysis of the specific experiments included in the PDG compilation.
 This is particularly so since the large CEX  cross section at resonance can
manifest in Bubble chambers as quickly decaying $K_S$'s.
 The same experiments could, however, ``bury" the whole notion of the $I=0$
 resonance if the $\Gamma_{\Theta^+}$ inferred turns out to be  too small.
 In this context we would like to add one final remark.

The color flux diagrams for the $ \bar{c}suud $ pentaquark depicted in Fig 21. (a) of the review paper on QCD inequalities\cite{LN}  can be viewed as merely mnemonics for the color couplings with each junction point  of three
 fluxes indicating the corresponding $\epsilon_{a,b,c}$ anti-symmetric color
coupling of the three $\mathbf{3}$ or $\mathbf{\bar{3}}$ representations to a color singlet.
  However, this figure may represent a true configuration-space picture of
 the pentaquark. In a strong coupling regime, we may wish to consider the flux lines as actual semiclassical configurations of flux tubes existing in configuration space. It is conceivable then that some tunnelling barrier
separates the $\Theta^+$ and the lower energy $K^{+}n$ baryon and meson state
obtained by annihilation (i.e., contraction of two $\epsilon-$ symbols) of
a junction and an anti-junction. In this case narrow widths could still
ensue, though it is not clear how narrow.

\section{Acknowledgements}

I have benefitted from discussions suggestions and comments of
   Alonso Botero, Boris Gelman, Terry Goldman, Vladimir Gudkov, Marek
   Karliner and Carl Rosenfeld.
   I am particularly indebted to  Ralph  Goethe for many long and very
   useful discussions and  David Tedeschi  for impressing
  upon  me  the weight of  evidence for the $\Theta^+$.

 \end{document}